\documentclass[prb,twocolumn,english,superscriptaddress,floatfix]{revtex4}
\usepackage{graphicx}
\usepackage{bm, amsmath, amssymb}
\usepackage{pdfpages}
\usepackage[T1]{fontenc}
\usepackage[latin9]{inputenc}
\usepackage{amstext}
\usepackage{esint}

\def\be{\begin{equation}}
\def\ee{\end{equation}}
\def\bea{\begin{eqnarray}}
\def\eea{\end{eqnarray}}

\usepackage{babel}

\newcommand{\rii}{\rho_{11}}

\newcommand{\roi}{\rho_{01}}

\newcommand{\add}[1]{{\color{black} #1}}

\begin{document}
\date{}
\title{}

%\begin{center}
%{\bf High dynamic range Josephson parametric amplifiers}\\
%K. W. Murch and N. Roch
%\end{center}

\title{Mapping quantum state dynamics in spontaneous emission}
%\title{Quantum state--measurement signal correlations in continuous weak measurement}

\author{M. Naghiloo}
\affiliation{Department of Physics, Washington University, St.\ Louis, Missouri 63130}
%\affiliation{Quantum Nanoelectronics Laboratory, Department of Physics, University of California, Berkeley CA 94720}
\author{N. Foroozani}
\affiliation{Department of Physics, Washington University, St.\ Louis, Missouri 63130}
\author{D. Tan}
\affiliation{Department of Physics, Washington University, St.\ Louis, Missouri 63130}
%\affiliation{Quantum Nanoelectronics Laboratory, Department of Physics, University of California, Berkeley CA 94720}
\author{A. Jadbabaie}
\affiliation{Department of Physics, Washington University, St.\ Louis, Missouri 63130}
\author{K. W. Murch*}
\affiliation{Department of Physics, Washington University, St.\ Louis, Missouri 63130}
\affiliation{Institute for Materials Science and Engineering, St.\ Louis, Missouri 63130}
\date{\today}

%\affiliation{Quantum Nanoelectronics Laboratory, Department of Physics, University of California, Berkeley CA 94720}
%\author{D}
%\affiliation{Department of Physics, Washington University, St.\ Louis, Missouri 63130}
%\affiliation{Quantum Nanoelectronics Laboratory, Department of Physics, University of California, Berkeley CA 94720}

%\affiliation{Department of Physics, Washington University, St.\ Louis, Missouri 63130}
\date{\today}

\begin{abstract}
The evolution of a quantum state undergoing radiative decay depends on how the emission is detected. We employ phase-sensitive amplification to perform homodyne detection of the spontaneous emission from a superconducting artificial atom. Using quantum state tomography, we characterize the correlation between the detected homodyne signal and the emitter's state, and map out the conditional back-action of homodyne measurement.  By tracking the diffusive quantum trajectories of the state as it decays, we characterize selective stochastic excitation induced by the choice of measurement basis. Our results demonstrate dramatic differences from the quantum jump evolution that is associated with photodetection and highlight how continuous field detection can be harnessed to control quantum evolution.

\end{abstract}

\maketitle

%Spontaneously emitted light is important in many fields of science ranging from fluorescence imaging to quantum encryption using single photons.  
  In spontaneous emission, an emitter decays from an excited state by releasing radiation into a quantized mode of the electromagnetic field.
 % added "excited state", changed "emission" to "releasing", felt like it was stale if we repeated "emitter" and "emission".
From the point of view of quantum measurement theory, the light-matter interaction entangles the quantum state of the emitter with its electromagnetic environment\cite{biln04,eich12}.
% This sounded a bit repetative, we just talked about "the quantized mode of the electromagnetic field", so I changed it to "environment"
 Subsequent measurements of the field convey information about the state of the emitter and consequently cause back-action \cite{wisebook}. 
 %"about and cause" is awkward. Reworded
Typically, spontaneous emission is detected in the form of energy quanta, resulting in an instantaneous jump of the emitter to a lower energy state.  However, if the emission is measured with a detector that is not sensitive to quanta, but rather to the amplitude of the field, the emitter's state undergoes different dynamics over finite timescales. Here, we use a near-quantum-limited Josephson parametric amplifier to perform continuous homodyne measurements of the spontaneous emission from a superconducting artificial atom. Under such detection, the emitter does not undergo jumps to its ground state, but rather diffuses through its state space. Furthermore, phase-sensitive operation of the amplifier squeezes the monitored field, inducing selective back-action on the emitter's state \cite{bolu14,wise12}.  
%Added line about phase-sensitive backaction. Best thing I could think of to quickly talk about y vs x back-action was "selective" back-action.  Anyway should it go here or one line earleir (before "under this detection")? 
% We use quantum state tomography and weak measurements to map out the conditional evolution and to track individual quantum trajectories \cite{wisebook,guer07,murc13traj} of the emitter's state, \add{showing how the choice of measurement on the field changes the quantum evolution of the emitter.  
\add{ Our results give insight into spontaneous emission and provide routes to control this light-matter interaction.  }
%Should also mention weak measurements, no?

%allowing fundamental tests of quantum mechanics \cite{Hens15}
%and in recent experiments heterodyne measurements have been used to observe the free decay of a superconducting qubit \cite{camp15}.

%These measurements are often considered in the context of resonance fluorescence \cite{Asta10} which has been used to observe  aspects of quantum evolution \cite{camp14}.

%The entanglement between a quantum emitter and its spontaneous emission field has been studied in experiments using natural atoms \cite{biln04} and in solid state systems \cite{eich12}, and can be used to herald entanglement between spatially separated systems \cite{bern13}. Spontaneous emission depends intimately on the fluctuations of the electromagnetic vacuum, and several experiments have controlled this process by either altering vacuum fluctuations \cite{murc13} or engineering the electromagnetic environment \cite{purc46,Houc07,Houc08,hoi15}. Emission can also be viewed in the context of quantum measurement; the light-matter interaction entangles the quantum state of the emitter with the quantized modes of the electromagnetic field \cite{wisebook}.  The entangled electromagnetic modes serve as a quantum pointer system; subsequent measurements of the field convey information about the emitter and cause back-action \cite{hatr13,groe13,murc13traj,camp15}.
%Therefore, the choice of measurement on the field can affect the evolution of the emitter's state \cite{wise12}. 

 \add{ Spontaneous emission depends intimately on the fluctuations of the electromagnetic vacuum, and several experiments have controlled this process by either altering vacuum fluctuations \cite{murc13} or engineering the electromagnetic environment \cite{purc46,Houc07,Houc08,hoi15}. The entanglement between a quantum emitter and its spontaneous emission field has been studied in experiments using natural atoms \cite{biln04} and solid state systems \cite{eich12}, and can be used to herald entanglement between spatially separated systems \cite{bern13}.  In the context of quantum measurement, the field can serve as a quantum pointer system \cite{wisebook}.   In this work we selectively measure a specific quadrature of this pointer system and map out the conditional evolution \cite{hatr13,groe13,murc13traj,camp15} of the emitter's state, showing how the choice of measurement on the field changes the conditional quantum evolution of the emitter. } %Therefore, the choice of measurement on the field can affect the evolution of the emitter's state \cite{wise12}.  

\begin{figure}
  \begin{center}
    \includegraphics[width=0.45\textwidth]{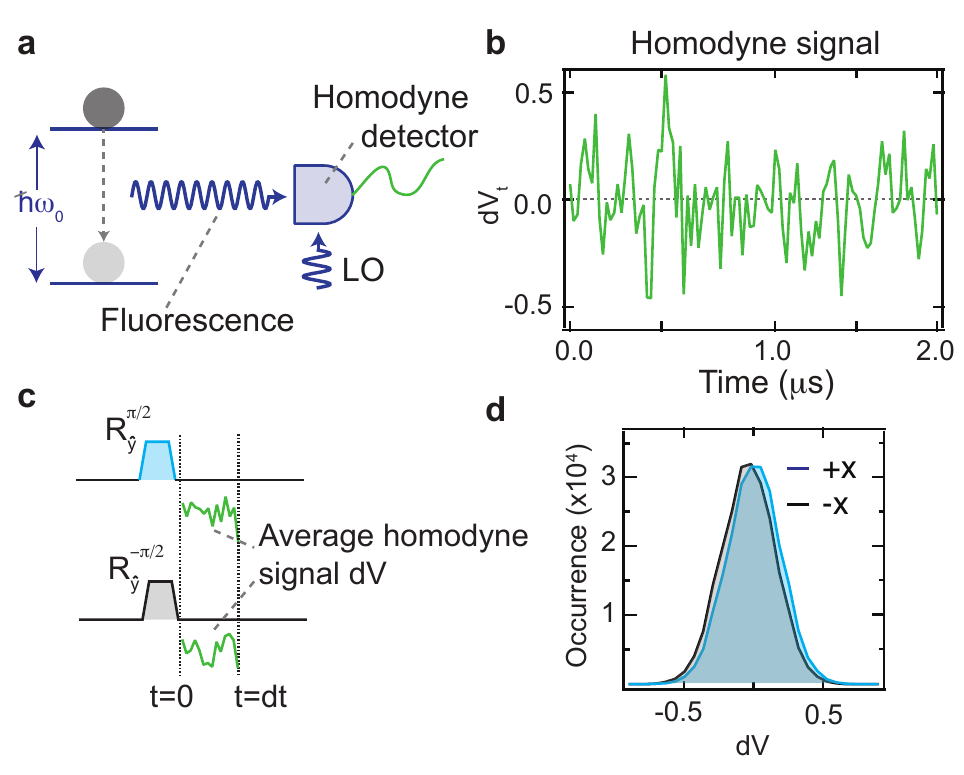}
  \end{center}
  \vspace{-.2in}
  \caption{\small {\bf Experimental setup.} {\bf a}, \add{The experiment uses a near-quantum-limited Josephson parametric amplifier to perform homodyne measurements of the fluorescence emitted by an effective two-level system.  {\bf b}, The dimensionless homodyne signal (denoted $dV_t$ at time step $t$) reflects the quantum fluctuations of the measured electromagnetic mode and is normalized so that its variance is $\gamma dt$. {\bf c} To calibrate the measurement, we prepare the emitter in the states $\pm x$ and average the ensuing homodyne signal for a time $dt = 20$ ns. {\bf d}, Histograms of the homodyne signals ({\bf c}) show how the measurement carries partial information about the $\sigma_x$ quadrature of the emitter's dipole.} }\label{fig1}
\end{figure}

 \begin{figure*}
\begin{center}
\includegraphics[width=0.8\textwidth]{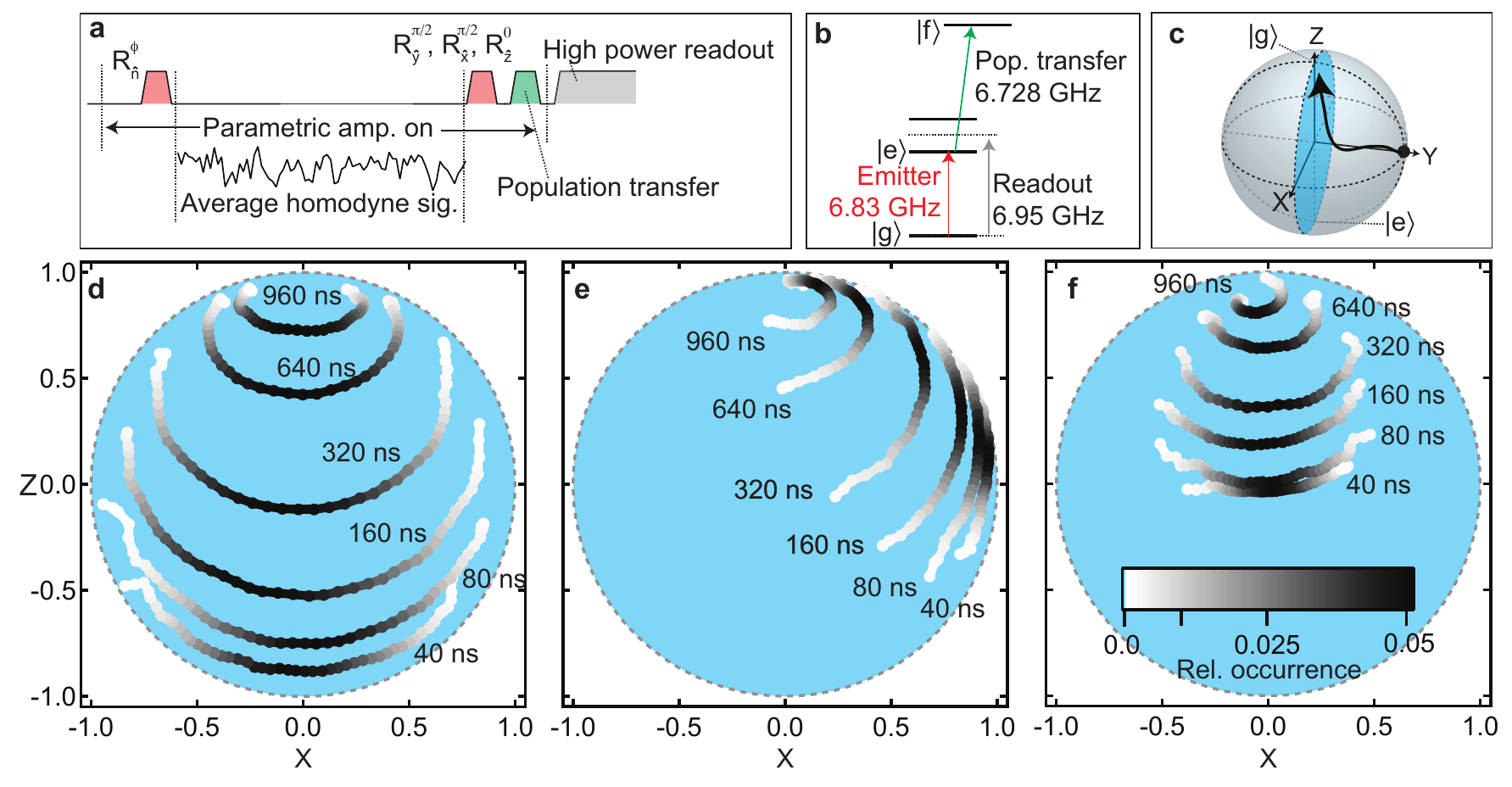}
\end{center}
\vspace{-.2in}
\caption{\small {\bf Mapping spontaneous decay.} {\bf a},  The experimental sequence prepares the emitter in an initial state and homodyne detection is used to record the emitted radiation.
Following a variable period of time, further rotations are applied to the emitter before state readout to perform quantum state tomography on the state. To enhance the readout contrast a pulse is applied to move the excited state population to a higher state of the system. {\bf b}, The level structure of the system and frequencies of the three microwave drives. {\bf c}, We average the state tomography to determine $x \equiv \langle \sigma_x \rangle |_{\bar{V}}$, and  $z\equiv \langle \sigma_z \rangle |_{\bar{V}}$ conditioned on the outcome of the homodyne measurement. These correlated tomography results are displayed on the $X$-$Z$ plane of the Bloch sphere for three different initial states: $-z$ ({\bf d}), $+x$ ({\bf e}), and $+y$ ({\bf f}). The color scale indicates the relative occurrence of the  of the each measurement value. Note the different backaction between ({\bf e}), and $+y$ ({\bf f}), a result of phase-sensitive amplification. 
} \label{fig2}
\end{figure*}

\add{Our system (Fig.\ 1a) consists of an effective two-level emitter formed by the resonant interaction of a transmon circuit\cite{koch07} and a three dimensional waveguide cavity\cite{paik113d}.  The strong light-matter interaction between the circuit and the cavity strips them of their individual character and gives rise to hybrid circuit-cavity states.} We use the lowest energy  transition ($\omega_0/2\pi = 6.83$ GHz) as an effective two-level system; deliberate coupling to a 50 $\Omega$ transmission line results in a radiative decay rate $\gamma = 2.3 \times 10^6\ $s$^{-1}$. The process of emission is described by the interaction Hamiltonian, $H_\mathrm{int} = \gamma (a^\dagger \sigma_- + a \sigma_+)$, where $a^\dagger\ (a)$ is the creation (annihilation) operator for a photon in the transmission line, and $\sigma_+ \ (\sigma_-)$ is the pseudo-spin raising (lowering) operator. \add{This interaction couples an arbitrary field quadrature $a^\dagger e^{i \phi} + a e^{-i \phi}$ to a corresponding emitter dipole $\sigma_-e^{i \phi} + \sigma_+e^{-i \phi}$. Due to the Heisenberg uncertainty relations, the outgoing radiation exhibits quantum fluctuations in its quadrature amplitudes. If these fluctuations are measured, they provide information on the emitter state and drive its stochastic evolution. %Conversely, if the fluctuations are de-amplified, their information is no longer available, eliminating the corresponding stochastic back-action on the emitter state.

To accurately detect these quantum fluctuations, we perform phase-sensitive amplification \cite{cler10} of outgoing signals near the emission frequency using a near-quantum-limited Josephson parametric amplifier \cite{cast08,hatr11para}. In this mode of operation, the amplifier squeezes the outgoing light along an axis in quadrature space given by the phase of the amplifier pump $\phi$. This constitutes a homodyne measurement of the amplified field quadrature $a^\dagger e^{i \phi} + a e^{-i \phi}$. Due to the emitter-field interaction, the choice of $\phi$ effectively enforces a choice of measurement basis on the emitter. In our experiment, we choose our amplifier phase $\phi=0$; the corresponding noisy homodyne signal (denoted $dV_t$, Fig.\ 1b) is then sensitive to the emitter dipole $\sigma_- + \sigma_+ = \sigma_x$.

The variance of the homodyne signal originates not only from the quantum fluctuations of the detected mode, but also from losses and added noise in the amplification chain. We account for this loss of information with the quantum efficiency $\eta$. The quantum noise is treated as a Weiner process; the fluctuations of the measurement signal $dV$ in an infinitesimal time step $dt$ are described by stochastic noise increments $dW_t$. Known as Weiner increments\cite{wisebook}, these are zero-mean, Gaussian random variables with variance $dt$. To accurately reflect this stochastic nature of the homodyne signal, we scale $dV$ such that it has a variance $\sigma^2 = \gamma dt$, with the full measurement record given by $dV_t =\sqrt{\eta}\gamma\langle\sigma_x\rangle dt + \sqrt{\gamma} dW_t$.

To experimentally demonstrate that our homodyne detection scheme is sensitive to a single quadrature of the emitter's dipole, we prepare the emitter in a specific state, perform homodyne measurement with $\phi=0$, and integrate the resulting signal (Fig.\ 1c,d). By repeating the measurement for several iterations, we can create histograms of the homodyne signal.  We compare the resulting distributions for two state preparations, $\pm x$ (the positive or negative eigenstates of the $\sigma_x$ Pauli operator). %We scale $dV$ such that its variance is $\sigma^2 =\gamma dt$, where $dt = 20$ ns is the integration time. 
The resulting separation of the two histograms, $\Delta V =2 \sqrt{\eta} \gamma  dt$, gives the quantum efficiency of our detection setup as $\eta = 0.3 $.}

\add{We now study the conditional dynamics of the emitter's state under radiative decay.  We conduct the experimental sequence depicted in Figure \ref{fig2}a; we first use a resonant rotation to prepare an initial state, then obtain the average homodyne signal $\bar{V}$ by integrating the detected homodyne signal for a variable period of time, and finally perform projective measurements to conduct quantum state tomography as described in the methods.} The results of these projective measurements are averaged conditionally on the integrated homodyne signal. This yields the conditional Pauli averages, $\langle \sigma_x \rangle |_{\bar{V}}$, $\langle \sigma_y \rangle |_{\bar{V}}$, $\langle \sigma_z \rangle |_{\bar{V}}$.  In Figure \ref{fig2}d-e we plot  $\langle \sigma_z \rangle |_{\bar{V}}$ and $\langle \sigma_x \rangle |_{\bar{V}}$ parametrically on the \add{$X$--$Z$} plane of the Bloch sphere for different integration times.  We study the conditional evolution for three different state preparations.

\add{When the emitter is prepared in the excited state (Fig.\ 2d), the $x$-component of the state develops a correlation with the average homodyne signal. }This highlights how our homodyne measurement provides an indirect signature \cite{jord15} of only the real part of $\sigma_- = (\sigma_x + i\sigma_y)/2$.  As the state is allowed more time to decay, it evolves to different deterministic arcs in the interior of the Bloch sphere. When the emitter is prepared in the state $+x$  (Fig.\ 2e), we observe that some of the conditioned states evolve toward the excited state \cite{bolu14}. This stochastic excitation is unique to amplitude measurements of the field quadrature, since such excitation is not possible under photodetection \cite{jord15}.  

%When the emitter is prepared in the excited state (along the $-z$ axis of the Bloch sphere, (Fig.\ 2d)), the $x$-component of the state develops a correlation with the average homodyne signal. With our choice of phase the measurement is only sensitive to the real part of $\sigma_- = (\sigma_x + i\sigma_y)/2$.  As the state is allowed more time to decay, it evolves to different deterministic arcs in the interior of the Bloch sphere. When the emitter is prepared in the state $+x$  (Fig.\ 2e), we observe that some of the conditioned states evolve toward the excited state \cite{bolu14}. This stochastic excitation is unique to amplitude measurements of the field quadrature, since such excitation is not possible under photo-detection \cite{jord15}. 

Under phase-sensitive amplification, the choice of homodyne phase can vary the stochastic back-action on the emitter's state.  To study this, we prepare the emitter in the state $+y$, an eigenstate of the imaginary part of our measured operator $\sigma_- = (\sigma_x + i\sigma_y)/2$. \add{This different state preparation is equivalent to preparing the emitter in the same state $+x$, (as depicted in Fig.\ 2e), and changing the homodyne phase by $\pi/2$. In this case, the emitter dipole  corresponds to the de-amplified quadrature of the emission field, and no stochastic excitation is observed (Fig.\ 2f). This demonstrates how the choice of homodyne measurement phase can be used to control the evolution of the emitter.}

\begin{figure}
  \begin{center}
    \includegraphics[width=0.5\textwidth]{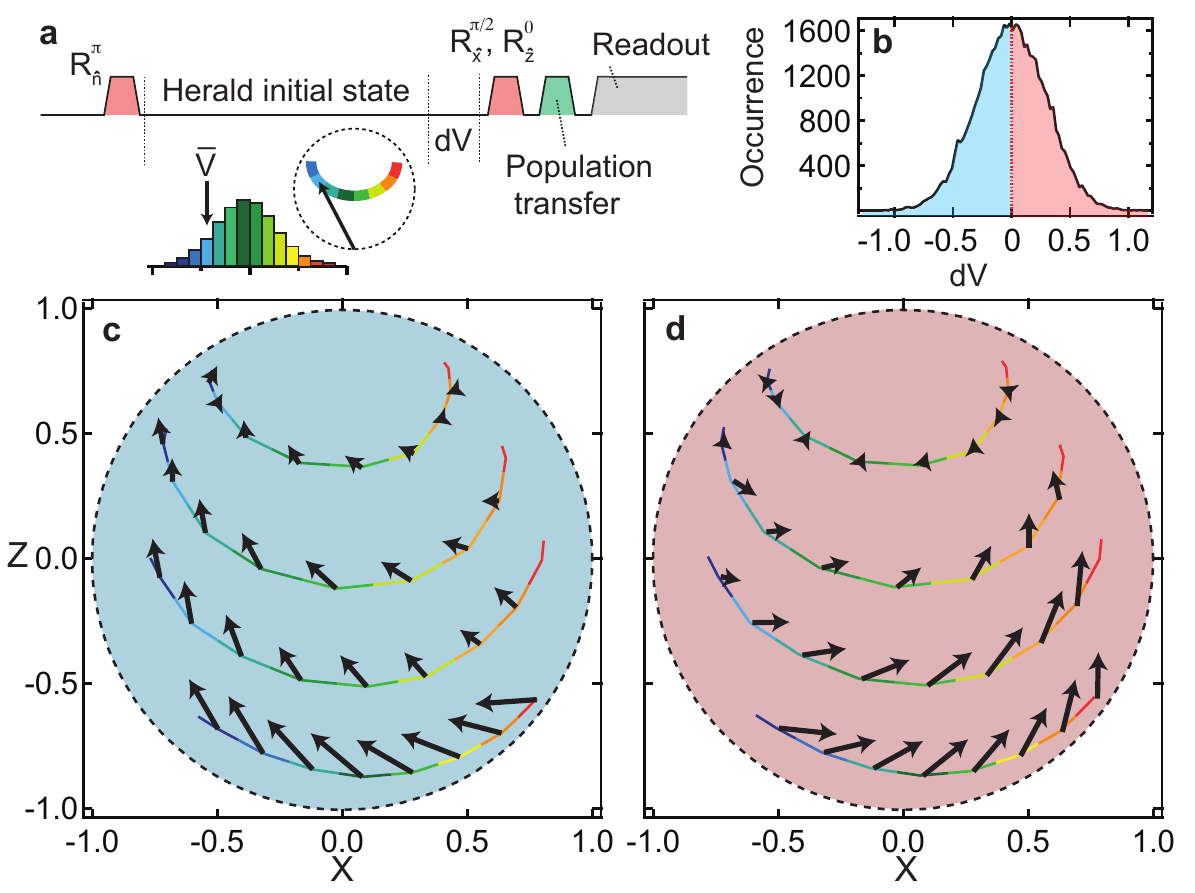}
  \end{center}
  \vspace{-.2in}
  \caption{\small {\bf Back-action vector maps.} {\bf a}, An arbitrary initial state in the \add{$X$-$Z$ plane} of the Bloch sphere is prepared by heralding on the average homodyne signal $\bar{V}$. Quantum state tomography is used to examine the conditional back-action based on a small portion of the signal $dV$. {\bf b}, Histogram of the signals $dV$ which we separate into positive or negative $dV$. The back-action imparted on the emitter for negative ({\bf c}) or positive ({\bf d}) values of $dV$ is depicted by an arrow at different locations in the \add{$X$-$Z$} plane of the Bloch sphere.   }\label{fig3}
\end{figure}

We take advantage of the deterministic evolution of the emitter, conditioned on the integrated homodyne signal, to characterize the back-action at different points in the Bloch sphere.  Figure \ref{fig3} shows a vector map of the state evolution due to a specific detected homodyne signal $dV$ at various points.  By preparing the emitter in the excited state and averaging the homodyne signal for various periods of time, we can prepare a nearly arbitrary mixed state through heralding.   After selecting a decay time and a specific initial state $(x_i, z_i)$,  based on an average signal $\bar{V}$, we digitize the homodyne signal for an additional time $dt=40$ ns to obtain $dV$. We then use quantum state tomography to determine the final state ($x_f, z_f$),  conditioned on the detection of $dV$ within a specified range.  The back-action at a specific location in state space, associated with the detection of a given value of $dV$, is provided by the vector connecting $(x_i, z_i)$ and  ($x_f, z_f$).  The back-action vector maps demonstrate how positive \add{(negative)} measurement results push the state toward $+x\add{(-x)}$. Furthermore, the maps show that the back-action is stronger near the state $-z$, indicating that the measurement strength is proportional \add{to} the emitter's excitation.
% Some minor changes here and there

%{\color{green} We take advantage of the deterministic evolution of the emitter to characterize the back-action at different points in the Bloch sphere.  Figure \ref{fig3} shows a vector map of the state evolution due to a specific detected homodyne signal $dV$ at various points. This graph which is obtain by theory independent method (see supplementary)  reveals that positive(negative) measurement is corresponding to a push of state toward the +x(-x) axis. It also shows that back-actions are stronger near the excited state. This results in the level of experiment suggests that the sign and magnitude of outcome signal can be used as measure for feedback to stabilize or steer the evolution hence controlling the dynamics of the decays}

The back-action maps that we present in Figure \ref{fig3} allow us to calculate the evolution of the emitter's state conditioned on a sequence of homodyne measurement results. Formally, this evolution is described by a stochastic master equation \cite{bolu14}, 
\begin{eqnarray}
d \rho = \gamma \mathcal{D}[\sigma_-] \rho dt+ \sqrt{\eta \gamma } \mathcal{H}[\sigma_- dW_t] \rho. \label{sme}
\end{eqnarray}
Where $ \mathcal{D}[\sigma_-] \rho = \sigma_- \rho \sigma_+ -\frac{1}{2} (\sigma_+ \sigma_- \rho + \rho  \sigma_+ \sigma_- )$ and $ \mathcal{H}[O]  \rho = O\rho + \rho O^\dagger - \mathrm{tr}[(O+O^\dagger)\rho]\rho$ are the dissipation and jump  superoperators, respectively.  When we ignore the results of homodyne monitoring (for example by setting $\eta=0$), the state follows deterministic evolution from an initial state to the ground state, as described by the first term of Eq. (\ref{sme}). \add{The second term accounts for information conveyed by the homodyne measurement through stochastic noise increments $dW_t$. %which are independent, zero-mean, Gaussian distributed Weiner increments with variance $dt$.  
% Our homodyne signal $dV_t$, detected in a time step $dt$, is proportional to the real part of $\sigma_-$ for the emitter, $dV_t = \sqrt{\eta }\gamma \langle \sigma_x \rangle dt  + \sqrt{\gamma}dW_t$.}
% Replaced "specific quadrature of the emitter's state" with mention of the real part of sigma_- 
We can recast this stochastic master equation in terms of the Bloch vector components $x,z,y$,}
\begin{eqnarray}
dx = - \frac{\gamma}{2} x dt + \sqrt{\eta} ( 1-z - x^2  )(dV_t - \gamma \sqrt{\eta} x dt), \label{smex}\\
dz =\gamma (1-z) dt + \sqrt{\eta} x(1-z)(dV_t - \gamma \sqrt{\eta}  x dt ), \label{smez}\\
dy = - \frac{\gamma}{2} y dt - \sqrt{\eta} xy (dV_t - \gamma \sqrt{\eta} x dt). \label{smey} 
 % I asked dian to double check that I change them by his result(mahdi)
\end{eqnarray}

\begin{figure}
  \begin{center}
    \includegraphics[width=0.5\textwidth]{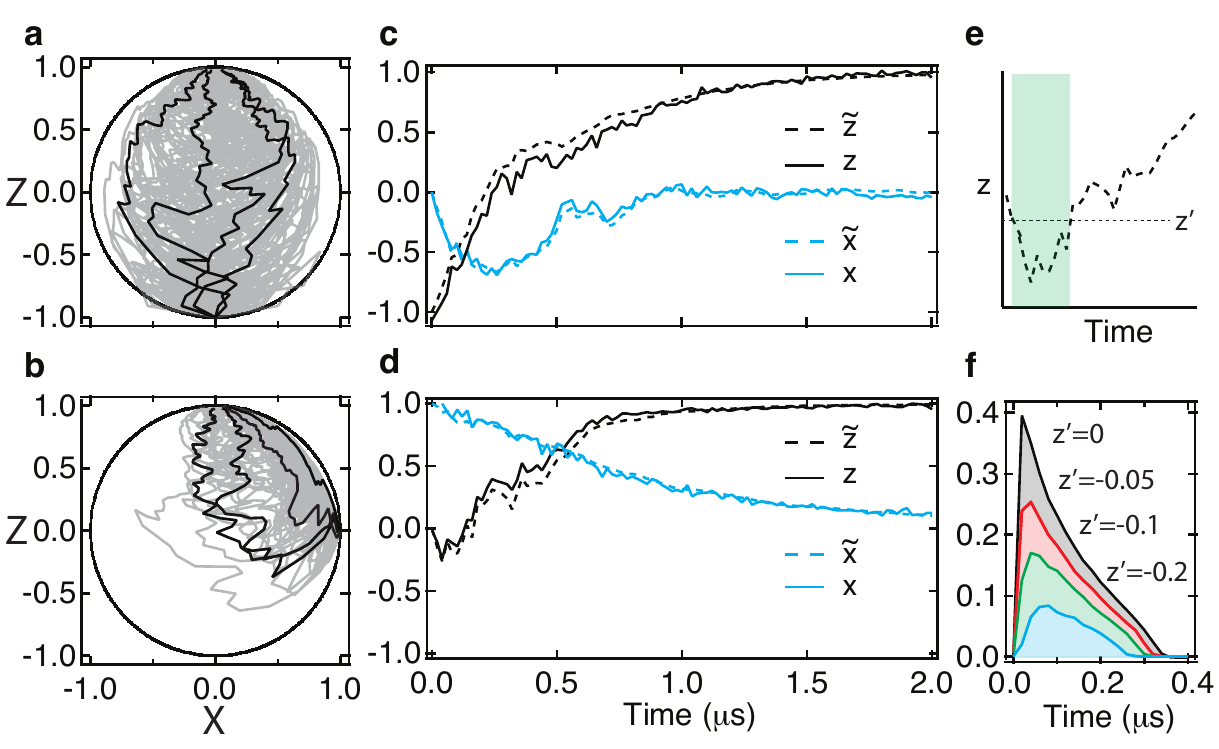}
  \end{center}
  \vspace{-.2in}
  \caption{\small {\bf Quantum trajectories.} {\bf a,b}, Quantum trajectories of spontaneous decay calculated by the stochastic master equation, initiated from $-z$ ({\bf a}) and $+x$ ({\bf b}).  Several trajectories are depicted in gray, and a few individual trajectories are highlighted in black. \add{{\bf c,d}  Individual trajectories ($\tilde{x}, \tilde{z}$) that originate from  $-z$ ({\bf c}) and $+x$ ({\bf d}) are shown as dashed lines and the tomographic reconstruction (see methods) based on projective measurements are shown as solid lines. }{\bf e}, For trajectories that are initiated along $+x$ some are excited (attaining values below a threshold $z'$).  {\bf f}, Fraction of the trajectories that are excited below the threshold $z'$ versus time. }\label{fig4}
\end{figure}

\add{We now turn to calculating individual quantum trajectories for the emitter's state as it evolves from an initial state.} In Figure \ref{fig4}, we prepare the emitter in the excited state and then digitize the detected homodyne signal for $2\ \mu$s.  Based on this signal, we use Eqs.\ (\ref{smex},\ref{smez},\ref{smey}) to calculate the emitter's trajectory using time steps of $dt = 20$ ns. \add{Instead of taking a straight path to the ground state, the trajectory diffuses through the Bloch sphere, subject to back-action from the measured quantum fluctuations of the emission field. } %As described in the methods, we use conditional quantum state tomography to verify that these individual trajectories make the correct predictions for all possible measurements at each time. %To accomplish this, we denote an individual trajectory $(\tilde{x}(t), \tilde{y}(t), \tilde{z}(t))$ (Note that $\tilde{y}(t)=0$).   At each time point, we perform several experiments of total duration $t'$, followed by one of three tomography and readout sequences. For each of these experiments, we calculate $(x(t'),z(t'))$; if $x(t')$ and $z(t')$ are within $\pm 0.12$ of $\tilde{x}(t')$ and  $\tilde{z}(t')$, then the subsequent tomography result is included in the tomographic validation at $t'$.  We follow this process for each $t'$ along the trajectory, resulting in a tomographic reconstruction of the trajectory.  
%As shown in Fig.\ \ref{fig4}, the tomographic validation is in good agreement with the individual trajectories, reproducing their specific stochastic behavior.
% Slight modifications

We also study quantum trajectories originating from the state $+x$. In this case, the stochastic back-action causes some of the trajectories to become more excited as they decay under homodyne detection. In Figure \ref{fig4}f, we quantify this feature by extracting the probability of excitation above a certain threshold at different times. By examining the measurement term in Eq.\ \ref{smez}, proportional to $\sqrt{\eta}$, we see that the state at $+x$ will be stochastically excited if the Weiner increment $dW_t$, obtained from the detected signal $dV_t$, is less than $- \sqrt{\gamma/\eta} dt$, predicting that $\sim 35\%$ of the trajectories should be excited in the first time step.
%This term is negative when the detected signal $dV_t$ is less than $\gamma\sqrt{\eta} dt$, the mean homodyne value when the emitter is in the state $+x$. If this stochastic kick is large enough to overcome the deterministic decay $\gamma dt$, the resulting back-action increases the excitation of the emitter, moving its state toward $-z$. %In terms of Weiner increments, the state at $+x$ will be stochastically excited if the measurement noise $dW_t$, obtained from the detected signal $dV_t$, is less than $ - \sqrt{\gamma/\eta} dt$.
%Adjusted it to reflect our discussion on Friday. Also I think I left out a square root when I was texting you then
%This process can also be understood in terms of information obtained about the emitter in the $\sigma_x$ basis; results where $dV_t<\gamma \sqrt{\eta} dt$ cause stochastic back-action that moves the emitter toward the state $-x$. However, because the state's evolution is confined to a deterministic arc in the Bloch sphere, this can result in excitation of the emitter. 

Recent experiments \cite{katz06,guer07,murc13traj,webe14,roch14,camp15} that harness Bayesian statistics or use quantum optics to track the evolution of quantum states have yielded a deeper understanding of quantum measurement evolution. \add{Here, we have shown how specific quadrature measurements of the fluorescence from a quantum emitter result in a rich conditional evolution of the state. We have harnessed this evolution to map out the back-action associated with such measurements, and we have tracked the individual quantum trajectories an emitter takes when decaying through fluorescence.} In contrast to the instantaneous dynamics of emission due to measurements of quanta, here we show that spontaneous emission may also occur over finite timescales.  %, with possible implications for further research in stochastic thermodynamics of decaying quantum systems.
%Cite thermo paper? Or no. Also new paragraph here or no?

Measurements, and more broadly, control over a quantum environment, can in principle be used to steer quantum evolution \cite{warr93,shap11}. Phase-sensitive parametric amplification squeezes the quantum pointer state and therefore causes selective measurement back-action on the emitter.
%Does squeezing need a citation?
%This gives an indication that our emitter is entangled with our Josephson amplifier, providing direction for further research into the quantum to classical transition for measurement signals.
% This means our measurement "collapse" is occuring at the amplifier
Such control over the quantum light-matter interaction has the potential to advance techniques in fluorescence based imaging, and will be essential in quantum feedback control \cite{wisebook,vija12,lang14} of quantum systems.

\vspace{.2in}

\vspace{.0 in}
{{\bf Methods}}

\vspace{.2 in}
{\bf Device fabrication and parameters}\\
  The emitter system consists of a transmon circuit characterized by charging energy $E_\mathrm{C}/h = 270$ MHz and Josephson energy $E_\mathrm{J}/h = 24.6$ GHz. The circuit was fabricated by double angle evaporation of aluminum on a high resistivity silicon substrate.  The circuit was then placed at the center of a waveguide cavity (dimensions $34.15 \times 27.9 \times 5.25$ mm) machined from 6061 aluminum.  The cavity geometry was chosen to be resonant with the lowest energy transition of the transmon circuit. The resonant interaction between the circuit and the cavity (characterized by coupling rate $g/2\pi = 136$ MHz) results in hybrid states, as described by the Jaynes-Cummings Hamiltonian. The cavity is deliberately coupled to 
 \add{two}  50 $\Omega$ cables: one weakly coupled port, characterized by coupling quality factor $Q_c \simeq 10^5$, is used to drive the system, while a more strongly coupled port $Q_c \simeq 10^4$  sets the total radiative decay time of the system.   This configuration results in an effectively ``one dimensional atom'', where all of the radiative decay is captured by the strongly coupled cable \cite{murc13}.  Spontaneous emission from this ``artificial atom'' is amplified by a near-quantum-limited Josephson parametric amplifier, consisting of a $1.5$ pF capacitor, shunted by a Superconducting Quantum Interference Device (SQUID) composed of two $I_0 = 1\ \mu$A Josephson junctions.  The amplifier is operated with negligible flux threading the SQUID loop and produces 20 dB of gain with an instantaneous 3-dB-bandwidth of 20 MHz. 

We used standard techniques to measure the energy decay time $T_1 = 430$ ns and Ramsey decay time $T_2^* = 830$ ns, indicating that the emitter experiences a negligibly small amount of pure dephasing.  We also examined the equilibrium state populations of the emitter using a Rabi driving technique \cite{geer13}, and found the excited state population to be $<3$\%.

{\bf State tracking} 

We use a master equation (equivalent to Eqs.\ (\ref{smex}-\ref{smey})) to propagate the density matrix for the emitter's state conditioned on the detected homodyne signal. The signal is digitized in 20 ns steps, and scaled such that its variance is $\gamma dt$.  At each time step, we update the density matrix components $\rii[i]$ and $\roi[i]$ based on the detected measurement signal $dV[i]$, where  \add{ $z\equiv 1-2 \rii$ } and $x\equiv 2 \mathrm{Re}[\roi]$.  \add{Our state update is consistent with the It\^o formulation of stochastic calculus.}
%\begin{eqnarray}
%\rii[i+1] =  \rii[i]-  \gamma\rii[i] dt  - \sqrt{\eta}(dV[i]-\sqrt{\eta} \gamma 2 \roi[i] dt)  (2\roi[i] \rii[i]),\\
%\roi[i+1]  	 =\roi[i]-\gamma \roi[i]/2 dt + \sqrt{\eta}(dV[i]-\sqrt{\eta}\gamma 2 \roi[i]dt)  (\rii[i]-2\roi[i] \roi[i]).
%\end{eqnarray}
%\begin{align}
%\rii[i+1] =  \rii[i]-  \gamma\rii[i] dt  - \sqrt{\eta}(dV[i]-\sqrt{\eta} \gamma 2 \roi[i] dt)  \times (2\roi[i] \rii[i])\\
%\roi[i+1]  	 =\roi[i]-\gamma \roi[i]/2 dt  + \sqrt{\eta}(dV[i]-\sqrt{\eta}\gamma 2 \roi[i]dt)  \times (\rii[i]-2\roi[i] \roi[i])
%\end{align}
\begin{align}
\rii[i+1] =  \rii[i&]-  \gamma\rii[i] dt \nonumber \\ &- \sqrt{\eta}(dV[i]-\sqrt{\eta} \gamma 2 \roi[i] dt) \\ &\quad \quad \quad \times (2\roi[i] \rii[i])\nonumber\\
\roi[i+1]  	 =\roi[i&]-\gamma \roi[i]/2 dt \nonumber \\ &+ \sqrt{\eta}(dV[i]-\sqrt{\eta}\gamma 2 \roi[i]dt) \\ &\quad \quad \quad \times (\rii[i]-2\roi[i] \roi[i])\nonumber
\end{align}

\begin{figure}
  \begin{center}
    \includegraphics[width=0.45\textwidth]{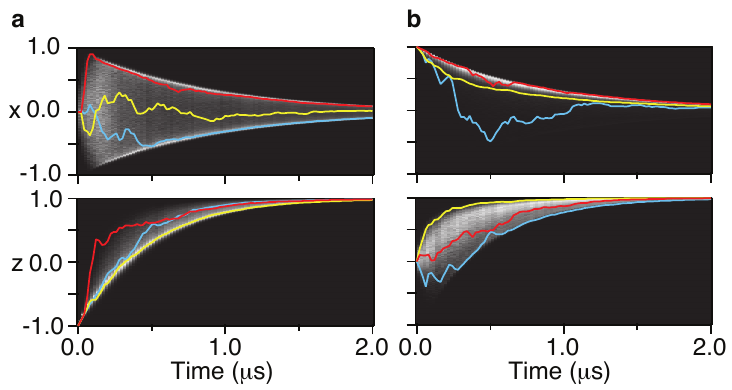}
  \end{center}
  \vspace{-.2in}
  \caption{\small {\bf State histograms.} Greyscale histograms represent the distribution for values of $x$ and $z$ at each time point. The greyscale shading is normalized such that the most frequent value is 1 at each time point. {\bf a}, Histograms of the state when the emitter is initialized in the state $-z$ with a few sample trajectories shown in color. {\bf b}, Histograms associated with decay from the excited state $+x$. }\label{histxz}
\end{figure}

\pagebreak

{\bf Ensemble dynamics}
\\
Based on \add{$9 \times 10^5$ } repetitions of the experiment and associated quantum trajectories, we can examine ensemble dynamics of the paths on the Bloch sphere taken by our decaying emitter. The behavior of single trajectories characterizes the dynamics of spontaneous decay subject to homodyne detection, and is distinctly different than the full ensemble behavior that decays deterministically toward the ground state.  

%\begin{figure}
%  \begin{center}
%   \includegraphics[width=0.5\textwidth]{accel2}
%  \end{center}
%  \vspace{-.2in}
%  \caption{\small {\bf Spontaneous decay dynamics.} Color maps of the ensemble average of the $x$-component of the  velocity ({\bf a}) and acceleration ({\bf b}) versus location of the emitter's state. }\label{velocitymap}
%\end{figure}

Figure \ref{histxz} displays greyscale histograms of the state at different points in time for two different initial conditions. For trajectories initialized in $-z$ (Fig.\ \ref{histxz}a), these histograms demonstrate how the decay paths are restricted to a deterministic arc in the Bloch sphere. Curiously enough, a state prepared in a traditional eigenstate of spontaneous emission will develop some quantum coherence when monitored under homodyne detection. \add{The $x$-components of such trajectories may be pinned to the edges of this arc on the $X$-axis, or instead may oscillate about the central value of $x=0$. We note that though the trajectories exhibit an immediate diffusive behavior for short timescales, the decay of coherence takes over at longer timescales, indicated by a decreasing upper bound on the stochastically acquired coherence. Examining behavior along the $Z$-axis, we see that though some trajectories may decay by more quickly approaching the ground state, no trajectory may decay more slowly in $z$ than a specific lower bound at each time step.}

On the other hand, when the emitter is initialized along $+x$ in a superposition of its excited and ground states, the histograms of the Bloch sphere coordinates show different behavior (Fig.\ \ref{histxz}b). The $x$-component of the trajectory encounters a decreasing upper bound on its maximum value, once more illustrating motion along a shrinking deterministic arc. The $z$-component, however, can exhibit extremely varied behavior. In addition to following the average decay path, the state may also stochastically excite, or it may rapidly decay in $z$ while approaching the surface of the Bloch sphere. Currently, it is these states that rapidly decay that have the highest purity on average, retaining the most information about the state. In comparison, due to our limited measurement efficiency, stochastically excited trajectories become more mixed as they diffuse toward the excited state. We note that for $\eta=1$, all of our trajectories, regardless of dynamics, would describe pure states confined to move only on the surface on the Bloch sphere. 

%Jump into deep end:

In fact, we expect the ensemble ratio of stochastically excited trajectories to increase with increasing $\eta$. As mentioned in the main text, trajectories experience $dz<0$ when the Weiner increment obtained from the measurement record satisfies $dW_t < -\sqrt{\gamma}dt/\sqrt{\eta}x$. Recall that $dW_t$ is a zero-mean random variable distributed with variance $dt$, and consider the back-action experienced by trajectories initialized with $x=1$. Naively, the probability of stochastic excitation is then given by the integral,
\begin{align}
 \int_{-\infty}^{-\sqrt{\gamma/\eta} dt} dW_t (2\pi dt)^{-1/2} e^{-dW_t^2/2dt}. \nonumber
\end{align}
As $\eta$ increases, so does the value of this integral. For $\eta=1$ and a time step $dt=20$ ns, the probability for spontaneous emission for our system reaches a maximum value of approximately 41.5\%. For our measured quantum efficiency of $\eta=0.3$, we expect approximately 35\% of trajectories to excite in the first time step.

\begin{figure}
  \begin{center}
    \includegraphics[width=0.5\textwidth]{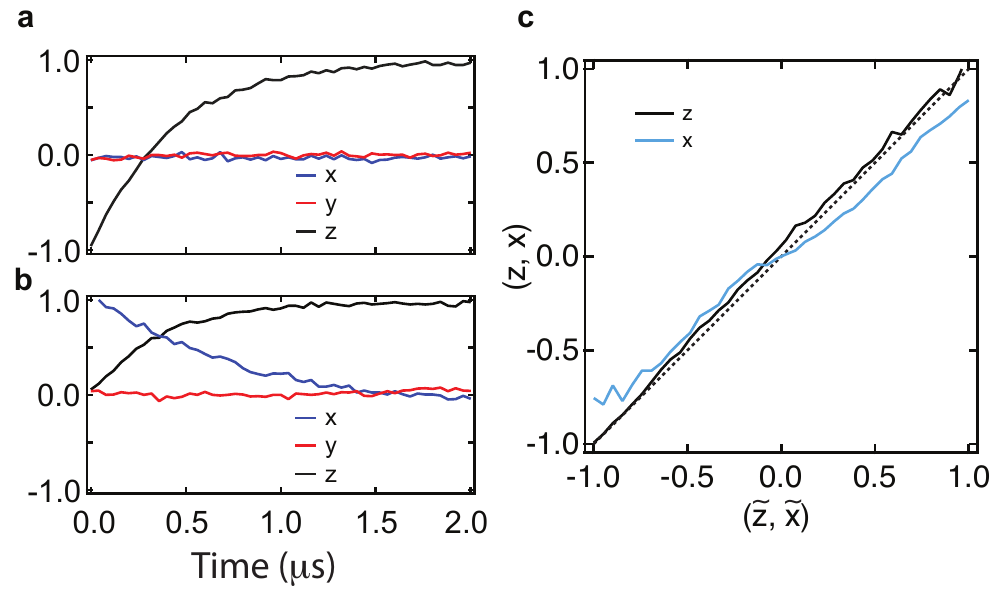}
  \end{center}
  \vspace{-.2in}
  \caption{\small {\bf Tomography calibrations.} The ensemble decay as determined by projective measurements for initial states $-z$ ({\bf a}), and $+x$ ({\bf b}). {\bf c}, Tomographic validation for the ensemble of trajectories shows the average tomography values $(x,z)$ versus the values obtained from individual trajectories. }\label{tomog}
\end{figure}

{\bf Tomography and readout calibration} 

%To perform quantum state tomography we apply either $\pi/2$ rotations about the $\hat{x}$ or $ \hat{y}$ axes, or perform no rotation, followed by projective measurements in the $\sigma_z$ basis. We use the Jaynes-Cummings nonlinearity technique \cite{reed10} to perform projective measurements. To enhance the readout contrast, we first apply a resonant pulse to transfer the excited state population to the $| f\rangle$ state (Fig.\ \ref{fig2}b).  

 All tomography results are corrected for imperfect state preparation and readout fidelities.  We perform state readout by first applying a resonant pulse at $6.73$ GHz to transfer the excited state population to a higher excited state, and then proceeding to drive the bare cavity resonance $6.95$ GHz at high power to conduct the Jaynes-Cummings high power readout technique \cite{reed10}.  Tomography for $y$ and $x$ is achieved by first applying a 40 ns $\pi/2$ rotation about the  $X$ or $Y$ axes.  The combined state preparation and readout fidelity (80\%) was determined from the contrast of resonant Rabi oscillations.   Each experimental sequence  includes separate calibration measurements used to determine the readout level of the ground state and the prepared excited state. These levels are used to scale the tomography results.  Figure \ref{tomog}a,b shows the ensemble decay curves for the state preparations $-z$ and $+x$.

The emitter's state is characterized by expectation values $(x, z)$.  To characterize accuracy of the state tracking, we compare the expectation values that are calculated for a single iteration of the experiment to the values obtained from an ensemble of projective measurements. In Figure \ref{fig4} we show this comparison to reconstruct and individual trajectory. 
 To accomplish this, we denote an individual trajectory $(\tilde{x}(t), \tilde{y}(t), \tilde{z}(t))$ (Note that $\tilde{y}(t)=0$).   At each time point, we perform several experiments of total duration $t'$, followed by one of three tomography and readout sequences. For each of these experiments, we calculate $(x(t'),z(t'))$; if $x(t')$ and $z(t')$ are within $\pm 0.12$ of $\tilde{x}(t')$ and  $\tilde{z}(t')$, then the subsequent tomography result is included in the tomographic validation at $t'$.  We follow this process for each $t'$ along the trajectory, resulting in a tomographic reconstruction of the trajectory. 

 We can further test the predictions given by the individual trajectories for all runs of the experiment at all times. Figure \ref{tomog}c displays the average projective measurement outcomes conditioned on the values of $\tilde{x}(t')$ or $\tilde{z}(t')$ compared to the values $\tilde{x}(t')$ or $\tilde{z}(t')$ showing good agreement between the individual trajectories and the projective measurements.

\begin{figure}
  \begin{center}
    \includegraphics[width=0.25\textwidth]{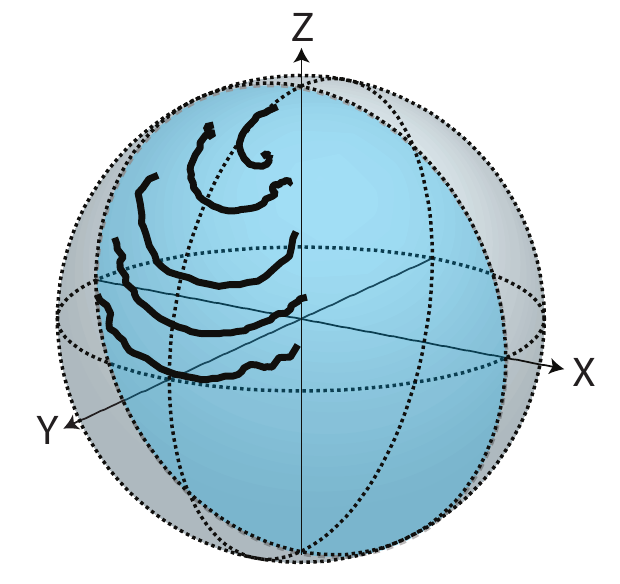}
  \end{center}
  \vspace{-.2in}
  \caption{\small {\bf Spontaneous decay from the state $+y$.} The emitter's state at different times conditioned on the integrated homodyne measurement signal. The decay times are  80, 160, 320, 640, 960 ns, and the data correspond to what is depicted in Figure \ref{fig2}f. The $x$--$z$ plane plotted in Figure \ref{fig2}f is highlighted in blue.  }\label{yprep3d}
\end{figure}

{\bf Phase-sensitive back-action} 

When the emitter is initialized in $+y$ the state dynamics are not confined to the $X$--$Z$ plane.  Figure \ref{yprep3d} displays the state conditioned on the integrated homodyne signal and shows how the $y$-component does not acquire a correlation with the measurement signal. This may be understood as a result of phase-sensitive amplification with $\phi=0$. When we perform our homodyne measurement of the real part of $\sigma_-$, we de-amplify the quadrature containing information on the imaginary part of $\sigma_-$, corresponding to $\sigma_y$ on the Bloch sphere. \add{The de-amplification of this orthogonal signal suppresses the magnitude of its quantum fluctuations, effectively eliminating the information associated with the $\sigma_y$ quadrature of the emitter's dipole.} Therefore we do not perform weak  measurements of $\sigma_y$, and we do not observe quantum dynamics such as stochastic excitation. 

%
%In this mode of operation, the amplifier also de-amplifies the orthogonal field quadrature $i(a^\dagger e^{i \phi} - a e^{-i \phi})$, significantly reducing back-action in the non-measured quadrature of the emitter's dipole, $i(\sigma_+e^{i \phi} - \sigma_-e^{-i \phi})$. 

We may also understand this phenomenon by examining the $dz$ and $dy$ segments of the stochastic master equation provided in the main text. The presence of an $xy$ coefficient on the measurement term in Eq.\ (\ref{smey}), means the stochastic back-action has no effect on the state when it is in an eigenstate of $\sigma_x$ or $\sigma_y$, limiting dynamics to a deterministic reduction in $y$. Meanwhile, if we examine Eq.\ (\ref{smez}) after factoring out a common factor of $(1-z)$ (which serves to push the trajectory toward the ground state) we see the measurement term is proportional only to $x$. Therefore, for a state prepared with $y=\pm 1$, there will be no initial stochastic excitation, and the state will begin its decay by deterministically approaching the ground state. However, once fluctuations in the measurement signal cause the state to acquire a nonzero $x$ value, the trajectory's dynamics will cease to be trivial.

\begin{figure*}
  \begin{center}
    \includegraphics[width=0.7\textwidth]{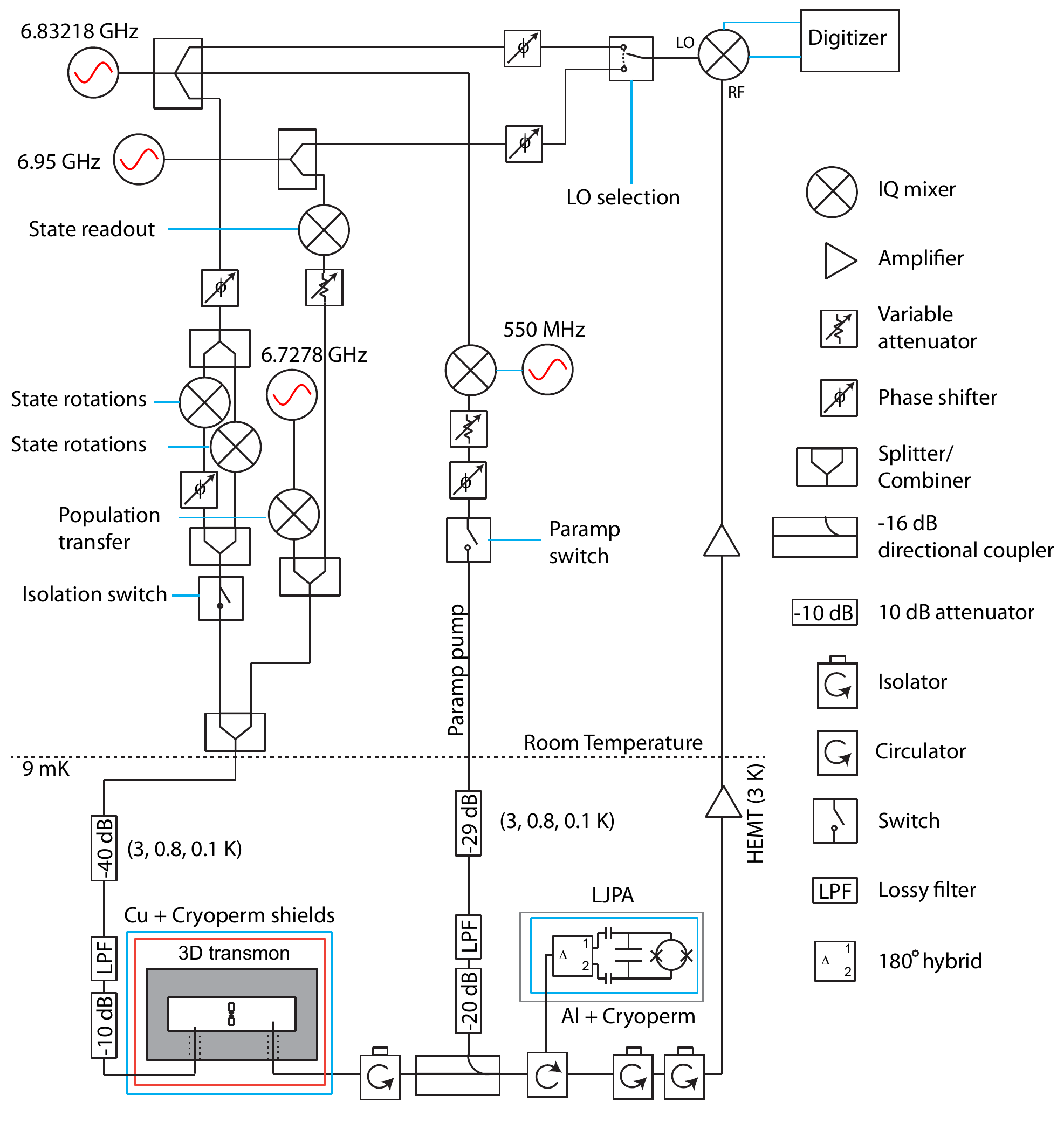}
  \end{center}
  \vspace{-.2in}
  \caption{\small {\bf Experimental setup.} }\label{schematic}
\end{figure*}

{\bf Experimental setup}

 Figure \ref{schematic} displays a simplified schematic of the experimental setup.  A single generator is used for qubit rotations, the amplifier pump, and demodulation of the amplified signal. The parametric amplifier is pumped by two sidebands that are equally separated from the carrier by 550 MHz, allowing for phase-sensitive amplification without leakage at the emitter's transition frequency.  The experimental repetition rate is $8$ kHz.

%\bibliographystyle{unsrt} %{apsrev}
%\bibliography{sponrefs}

\begin{thebibliography}{10}

\bibitem{biln04}
B.~B. Blinov, D.~L. Moehring, L.-M. Duan, and C.~Monroe.
\newblock Observation of entanglement between a single trapped atom and a
  single photon.
\newblock {\em Nature}, 428:153--157, 2004.

\bibitem{eich12}
C.~Eichler, C.~Lang, J.~M. Fink, J.~Govenius, S.~Filipp, and A.~Wallraff.
\newblock Observation of entanglement between itinerant microwave photons and a
  superconducting qubit.
\newblock {\em Phys. Rev. Lett.}, 109:240501, Dec 2012.

\bibitem{wisebook}
H.~Wiseman and G.~Milburn.
\newblock {\em Quantum Measurement and Control}.
\newblock Cambridge University Press, 2010.

\bibitem{bolu14}
Anders Bolund and Klaus M\o{}lmer.
\newblock Stochastic excitation during the decay of a two-level emitter subject
  to homodyne and heterodyne detection.
\newblock {\em Phys. Rev. A}, 89:023827, Feb 2014.

\bibitem{wise12}
Howard~M. Wiseman and Jay~M. Gambetta.
\newblock Are dynamical quantum jumps detector dependent?
\newblock {\em Phys. Rev. Lett.}, 108:220402, May 2012.

\bibitem{murc13}
K.~W. Murch, S.~J. Weber, K.~M. Beck, E.~Ginossar, and I.~Siddiqi.
\newblock Reduction of the radiative decay of atomic coherence in squeezed
  vacuum.
\newblock {\em Nature}, 499:62--65, 2013.

\bibitem{purc46}
E.~M. Purcell.
\newblock Spontaneous emission probabilities at radio frequencies.
\newblock {\em Phys. Rev.}, 69:681, 1946.

\bibitem{Houc07}
A.~A. Houck, D.~I. Schuster, J.~M. Gambetta, J.~A. Schreier, B.~R. Johnson,
  J.~M. Chow, L.~Frunzio, J.~Majer, M.~H. Devoret, S.~M. Girvin, and R.~J.
  Schoelkopf.
\newblock Generating single microwave photons in a circuit.
\newblock {\em Nature}, 449:328--331, 2007.

\bibitem{Houc08}
A.~A. Houck, J.~A. Schreier, B.~R. Johnson, J.~M. Chow, Jens Koch, J.~M.
  Gambetta, D.~I. Schuster, L.~Frunzio, M.~H. Devoret, S.~M. Girvin, and R.~J.
  Schoelkopf.
\newblock Controlling the spontaneous emission of a superconducting transmon
  qubit.
\newblock {\em Phys. Rev. Lett.}, 101:080502, Aug 2008.

\bibitem{hoi15}
I.-C. Hoi, A.~F. Kockum, L.~Tornberg, A.~Pourkabirian, G.~Johansson,
  P.~Delsing, and C.~M. Wilson.
\newblock Probing the quantum vacuum with an artificial atom in front of a
  mirror.
\newblock {\em Nature Physics}, page 3484, 2015.

\bibitem{bern13}
H.~Bernien, B.~Hensen, W.~Pfaff, G.~Koolstra, M.~S. Blok, L.~Robledo, T.~H.
  Taminiau, M.~Markham, D.~J. Twitchen, L.~Childress, and R.~Hanson.
\newblock Heralded entanglement between solid-state qubits separated by three
  metres.
\newblock {\em Nature}, 497:86--90, 2013.

\bibitem{hatr13}
M.~Hatridge, S.~Shankar, M.~Mirrahimi, F.~Schackert, K.~Geerlings, T.~Brecht,
  K.~M. Sliwa, B.~Abdo, L.~Frunzio, S.~M. Girvin, R.~J. Schoelkopf, and M.~H.
  Devoret.
\newblock Quantum back-action of an individual variable-strength measurement.
\newblock {\em Science}, 339(6116):178--181, 2013.

\bibitem{groe13}
J.~P. Groen, D.~Rist\`e, L.~Tornberg, J.~Cramer, P.~C. de~Groot, T.~Picot,
  G.~Johansson, and L.~DiCarlo.
\newblock Partial-measurement backaction and nonclassical weak values in a
  superconducting circuit.
\newblock {\em Phys. Rev. Lett.}, 111:090506, Aug 2013.

\bibitem{murc13traj}
K.~W. Murch, S.~J. Weber, C.~Macklin, and I.~Siddiqi.
\newblock Observing single quantum trajectories of a superconducting qubit.
\newblock {\em Nature}, 502:211, 2013.

\bibitem{camp15}
P.~Campagne-Ibarcq, P.~Six, L.~Bretheau, A.~Sarlette, M.~Mirrahimi, P.~Rouchon,
  and B.~Huard.
\newblock Observing quantum state diffusion by heterodyne detection of
  fluorescence.
\newblock {\em Phys. Rev. X}, 6:011002, Jan 2016.

\bibitem{koch07}
Jens Koch, Terri~M. Yu, Jay Gambetta, A.~A. Houck, D.~I. Schuster, J.~Majer,
  Alexandre Blais, M.~H. Devoret, S.~M. Girvin, and R.~J. Schoelkopf.
\newblock Charge-insensitive qubit design derived from the cooper pair box.
\newblock {\em Phys. Rev. A}, 76:042319, Oct 2007.

\bibitem{paik113d}
Hanhee Paik, D.~I. Schuster, Lev~S. Bishop, G.~Kirchmair, G.~Catelani, A.~P.
  Sears, B.~R. Johnson, M.~J. Reagor, L.~Frunzio, L.~I. Glazman, S.~M. Girvin,
  M.~H. Devoret, and R.~J. Schoelkopf.
\newblock Observation of high coherence in josephson junction qubits measured
  in a three-dimensional circuit qed architecture.
\newblock {\em Phys. Rev. Lett.}, 107:240501, Dec 2011.

\bibitem{cler10}
A.~A. Clerk, M.~H. Devoret, S.~M. Girvin, Florian Marquardt, and R.~J.
  Schoelkopf.
\newblock Introduction to quantum noise, measurement, and amplification.
\newblock {\em Rev. Mod. Phys.}, 82:1155--1208, Apr 2010.

\bibitem{cast08}
M.~A. Castellanos-Beltran, K.~D. Irwin, G.~C. Hilton, L.~R. Vale, and K.~W.
  Lehnert.
\newblock Amplification and squeezing of quantum noise with a tunable josephson
  metamaterial.
\newblock {\em Nature Physics}, 4:929--931, 2008.

\bibitem{hatr11para}
M.~Hatridge, R.~Vijay, D.~H. Slichter, John Clarke, and I.~Siddiqi.
\newblock Dispersive magnetometry with a quantum limited squid parametric
  amplifier.
\newblock {\em Phys. Rev. B}, 83:134501, Apr 2011.

\bibitem{jord15}
Andrew~N. Jordan, Areeya Chantasri, Pierre Rouchon, and Benjamin Huard.
\newblock Anatomy of fluorescence: Quantum trajectory statistics from
  continuously measuring spontaneous emission.
\newblock {\em arXiv:1511.06677}, 2015.

\bibitem{katz06}
N.~Katz, M.~Ansmann, Radoslaw~C. Bialczak, Erik Lucero, R.~McDermott, Matthew
  Neeley, Matthias Steffen, E.~M. Weig, A.~N. Cleland, John~M. Martinis, and
  A.~N. Korotkov.
\newblock Coherent state evolution in a superconducting qubit from
  partial-collapse measurement.
\newblock {\em Science}, 312(5779):1498--1500, 2006.

\bibitem{guer07}
C.~Guerlin, J.~Bernu, S.~Deleglise, C.~Sayrin, S.~Gleyzes, S.~Kuhr, M.~Brune,
  J.~Raimond, and S.~Haroche.
\newblock Progressive field-state collapse and quantum non-demolition photon
  counting.
\newblock {\em Nature}, 448:889, 2007.

\bibitem{webe14}
S.~J. Weber, A.~Chantasri, J.~Dressel, A.~N. Jordan, K.~W. Murch, , and
  I.~Siddiqi.
\newblock Mapping the optimal route between two quantum states.
\newblock {\em Nature}, 511:570Ð573, 2014.

\bibitem{roch14}
N.~Roch, E.~Schwartz, M.\, F.~Motzoi, C.~Macklin, R.~Vijay, W.~Eddins, A.\,
  N.~Korotkov, A.\, B.~Whaley, K.\, M.~Sarovar, and I.~Siddiqi.
\newblock Observation of measurement-induced entanglement and quantum
  trajectories of remote superconducting qubits.
\newblock {\em Phys. Rev. Lett.}, 112:170501, Apr 2014.

\bibitem{warr93}
W.S. Warren, H.~Rabitz, and M.~Dahleh.
\newblock Coherent control of quantum dynamics: The dream is alive.
\newblock {\em Science}, 259:1581--1589, 1993.

\bibitem{shap11}
Moshe Shapiro and Paul Brumer.
\newblock {\em Quantum Control of Molecular Processes}.
\newblock Wiley-VCH Verlag GmbH \& Co. KGaA, 2011.

\bibitem{vija12}
R.~Vijay, C.~Macklin, D.~H. Slichter, S.~J. Weber, K.~W. Murch, R.~Naik, A.~N.
  Korotkov, and I.~Siddiqi.
\newblock Stabilizing rabi oscillations in a superconducting qubit using
  quantum feedback.
\newblock {\em Nature}, 490:77, 2012.

\bibitem{lang14}
G.~de~Lange, D.~Rist\`e, M.~J. Tiggelman, C.~Eichler, L.~Tornberg,
  G.~Johansson, A.~Wallraff, R.~N. Schouten, and L.~DiCarlo.
\newblock Reversing quantum trajectories with analog feedback.
\newblock {\em Phys. Rev. Lett.}, 112:080501, Feb 2014.

\bibitem{geer13}
K.~Geerlings, Z.~Leghtas, I.~M. Pop, S.~Shankar, L.~Frunzio, R.~J. Schoelkopf,
  M.~Mirrahimi, and M.~H. Devoret.
\newblock Demonstrating a driven reset protocol for a superconducting qubit.
\newblock {\em Phys. Rev. Lett.}, 110:120501, Mar 2013.

\bibitem{reed10}
M.~D. Reed, L.~DiCarlo, B.~R. Johnson, L.~Sun, D.~I. Schuster, L.~Frunzio, and
  R.~J. Schoelkopf.
\newblock High-fidelity readout in circuit quantum electrodynamics using the
  jaynes-cummings nonlinearity.
\newblock {\em Phys. Rev. Lett.}, 105:173601, Oct 2010.

\end{thebibliography}

{\bf Acknowledgements}  We thank A. N. Jordan and K. M\o lmer for discussions. This research was supported in part by the John F. Templeton Foundation and the Sloan Foundation and used facilities at the Institute of Materials Science and Engineering at Washington University.

%{\bf Author contributions}
%M. N., N. F.  and K.W.M. performed the experiments and D. T. fabricated the samples. All authors contributed to the data analysis and writing the manuscript.
%
%{\bf Author information} 
% Reprints and permissions information is available at www.nature.com/reprints.
% 
% The authors declare no competing financial interests. 
% 
Correspondence and requests for materials should be addressed to K.W.M. (murch@physics.wustl.edu)

%These state update equations encompass the back-action vector map that we experimentally probe in Fig.\ \ref{fig3}.

%{\bf Author contributions}
%M. N., N. F.  and K.W.M. performed the experiments and D. T. fabricated the samples. All authors contributed to the data analysis and writing the manuscript.

%{\bf Author information} 
 %Reprints and permissions information is available at www.nature.com/reprints.
 
 %The authors declare no competing financial interests. 

\pagebreak

%%%%% FIG1

%%%% FIG2

%%% FIG3

%%%%FIG4

%%%%% EXTENDED DATA

\end{document}